\documentclass[aps,10pt,twoside,tightenlines,superscriptaddress]{revtex4}
\usepackage{amsmath,amsfonts,latexsym,graphicx}
\usepackage[colorlinks=false]{hyperref} 

\newcommand{\ep}{\epsilon}
\newcommand{\ket}[1]{\left| #1\right\rangle}        
\newcommand{\bra}[1]{\left\langle #1\right|}        
\newcommand{\kets}[1]{|#1\rangle}        
\newcommand{\half}{\frac{1}{2}}
\newcommand{\ii}{\mathbb{I}}

\newcommand{\norm}[1]{\left\| #1\right\|}        

\begin{document}


\title{A new construction for a QMA complete 3-local Hamiltonian.}
\author{Daniel Nagaj}\email{nagaj@mit.edu}
\affiliation{Center for Theoretical Physics, MIT, Cambridge, MA 02139} 
\author{Shay Mozes}
\email{shaymozes@gmail.com}
\thanks{Work conducted while visiting MIT}
\noaffiliation
\date{December 12, 2006}

\begin{abstract}

We present a new way of encoding a quantum computation into 
a 3-local Hamiltonian.
Our construction is novel in that it does not include any terms that
induce legal-illegal clock transitions. Therefore, the weights of the
terms in the Hamiltonian do not scale with the size of the problem as in previous constructions.
This improves the construction by Kempe and Regev \cite{KR03}, who were the first 
to prove that 3-local Hamiltonian is complete for the complexity class QMA, the quantum 
analogue of NP.

Quantum k-SAT, a restricted version of the local Hamiltonian problem using only projector terms, 
was introduced by Bravyi \cite{Bravyi} as an analogue of the classical k-SAT problem.
Bravyi proved that quantum 4-SAT is complete for the class QMA with
one-sided error (QMA$_1$) and that quantum
2-SAT is in P. 
We give an encoding of a quantum circuit into 
a quantum 4-SAT Hamiltonian using only 3-local terms. 
As an intermediate step to this 3-local construction, 
we show that quantum 3-SAT for particles with dimensions $3\times 2\times 2$ 
(a qutrit and two qubits) is QMA$_1$ complete.
The complexity of quantum 3-SAT with qubits remains an open question.

\end{abstract}
\maketitle


\section{Introduction}

In recent years, quantum complexity classes have been defined and studied in an attempt to understand
the capacity and limitations of quantum computers and quantum algorithms and their relation to 
classical complexity classes. The complexity class QMA, also known as BQNP, was studied and 
defined in \cite{Kni96} and in \cite{Kitaevbook} as the quantum analogue of the classical complexity 
class NP in a probabilistic setting. The local-Hamiltonian problem, a quantum analogue of classical 
satisfiability problems such as MAX-k-SAT, is an example of a complete problem for the class QMA.
Initially, building on ideas that go back to Feynman \cite{Feynman}, Kitaev \cite{Kitaevbook} 
has shown that 5-local Hamiltonian is QMA complete. 
Later, Kempe and Regev \cite{KR03} showed that 3-local Hamiltonian is QMA complete.
Using perturbation theory gadgets, this result was further improved 
to show that 2-local Hamiltonian is QMA complete as well \cite{KKR04},\cite{Terhal2D}.
The basic ingredient in all of these proofs is a reduction between quantum circuits and time independent
local Hamiltonians. 

Recently, quantum $k$-SAT, the special case where the Hamiltonian is a sum of local projectors was defined 
and studied by Bravyi \cite{Bravyi} as a natural analogue of classical $k$-SAT. There it was shown that 
quantum 2-SAT is in P and that quantum 4-SAT is complete for QMA$_1$ (QMA with single sided error). The classification
of quantum 3-SAT is still an open question.

In this work we show a new reduction from a verifier quantum circuit to a 3-local Hamiltonian. The novelty
of our construction is that it leaves the space of legal clock-register states invariant.
Therefore, the weights of the terms in our Hamiltonian do not scale with the size of of the input problem. 
Such terms do appear in the constructions of \cite{KR03} and \cite{KKR04}.
As an intermediate step in our construction, we prove that quantum 3-SAT for qutrits is QMA$_1$-complete.

The paper is organized as follows. We review the necessary background in Section \ref{prel}.
In Section \ref{3sat} we present a qutrit-clock construction and show that quantum 3-SAT for particles with
dimensions $3 \times 2 \times 2$ (the interaction terms in the Hamiltonian couple one qutrit and two qubits) 
is QMA$_1$-complete. 
The existence of such a construction was previously mentioned but not specified 
by Bravyi and DiVincenzo in \cite{Bravyi} as \cite{DiVincenzo}.
In Section \ref{3loc} we show how to encode
the qutrit clock particles from Section \ref{3sat} into a pair of qubits in such 
a way that the Hamiltonian remains 3-local. Thus we obtain
a new construction of a QMA complete 3-local Hamiltonian.
This Hamiltonian is composed of 4-local positive operator terms.
However, each of these 4-local positive operators is composed of 
only 3-local interaction terms.
This is not a quantum 3-SAT Hamiltonian, since the 3-local terms by themselves 
are not positive operators. 
We discuss the complexity of quantum 3-SAT and further directions in Section \ref{conclusions}.


\section{Preliminaries} \label{prel}

\subsection{The class QMA}
A promise problem $L=\{L_{yes} \cup L_{no}\}$ of size $N$ is in the class QMA
if there exists a polynomial time quantum verifier circuit $V$ such that 
\begin{enumerate}
	\item $\forall x\in L_{yes}$ : there exists a witness state $\ket{\varphi(x)}$ such that 
		the computation $V\ket{x} \otimes \ket{\varphi(x)} \otimes \ket{0\dots0}_{ancilla}$
		yields the answer 1 with probability at least $p$ ;
	\item $\forall x\in L_{no}$ : for any witness state $\ket{\varphi(x)}$, the computation 
		$V\ket{x} \otimes \ket{\varphi(x)} \otimes \ket{0\dots0}_{ancilla}$
		yields the answer 1 with probability at most $p-\epsilon$,
\end{enumerate}
where $p>0$ and $\epsilon = \Omega(1/\textrm{poly}(N))$.
The class QMA$_1$ is the class QMA with single sided error, i.e. with $p=1$ in the above definition.
Throughout this paper, we will use the notation
$V \ket{x} \otimes \ket{\varphi(x)} \otimes \ket{0\dots0} = U \ket{\varphi} \otimes \ket{0\dots0}$,
meaning that the verifier circuit $U$ is the verifier circuit $V$ for the specific instance $x$ 
of the problem $L$.


\subsection{The local-Hamiltonian problem}
\label{lochamsection}
An operator $H$ acting on $N$ qubits is said to be $k$-local if $H$ can be expressed as the sum of
Hermitian operators, each acting on at most $k$ qubits.
In the \textsc{local-Hamiltonian} decision problem, we are given 
a description of a $k$-local Hamiltonian on $N$ qubits, $H=\sum_{j=1}^r H_j$ with $r=\textrm{poly}(N)$.
Each $H_j$ has a bounded operator norm $||H_j|| \leq \textrm{poly}(N)$ and its entries are specified by
poly$(N)$ bits. In addition, we are given two constants $a<b$ with $b-a=\Omega(1/\textrm{poly}(N))$.
We have to decide whether the ground state energy of $H$ is at most $a$ (``yes'' instance) or at least $b$ 
(``no'' instance).

It was shown in \cite{Kitaevbook} that the $k$-local Hamiltonian problem for any constant $k$ is in QMA.
Therefore, in order to prove that $k$-local Hamiltonian is QMA complete, we now need to
show that given a quantum verifier circuit $U$ and constants $p,\epsilon$ as in the definition of QMA,
we can construct a $k$-local Hamiltonian $H$ and find constants $a$ and $b$ in polynomial time, 
with the following properties. 
If $\exists \ket{\varphi}$ such that 
the computation $U\ket{\varphi} \otimes \ket{0\dots0}$ yields the answer 
1 with probability at least $p$, the groundstate energy of $H$ has is at most $a$. 
On the other hand, if $\forall \ket{\varphi}$, the computation 
$U\ket{\varphi} \otimes \ket{0\dots0}$ yields the answer 1 with probability at most $p-\epsilon$, 
the ground state energy of $H$ is greater than $b$. 

The idea of encoding a unitary computation into the ground state of a time independent Hamiltonian 
goes back to Feynman \cite{Feynman}. 
To encode the quantum computation $U = U_L \dots U_2 U_1$ with $L$ steps on an unknown $N$-qubit 
input $\ket{\varphi}$ (and $N_a$ ancilla qubits) into the ground state of a Hamiltonian, we define 
a Hamiltonian acting on the space of $N+N_a$ work qubits, and on a clock register.
We represent the computational history for the computation $U$ by the state 
$\ket{\phi} = \sum \ket{\phi_k}_{work} \otimes \ket{C_k}_{clock}$, where 
the states $\ket{C_k}_{clock}$ for $1 \leq k \leq L+1$ are $L+1$ orthogonal legal states of the clock 
register representing time $1 \leq k \leq L+1$ and the state 
$\ket{\phi_k}_{work}$ represents the state of the work qubits at time $k$.
The expectation value of this Hamiltonian in a quantum state $\ket{\psi}$ ``checks'' whether the state encodes
a valid quantum computation that yielded a ``yes'' answer. This is achieved by constructing the Hamiltonian as a sum
of positive terms that penalize (i.e., increase the energy of) states that do not encode a legal computation,
and terms that penalize legal computational history states that yielded the ``no'' answer:
\begin{eqnarray}
	H &=& H_{clock} + H_{init} + H_{out} + \sum_{k=1}^{L} H_{prop}^{(k)}. \label{ham5}
\end{eqnarray}
The term $H_{clock}$ acts only on the clock register and penalizes illegal states of the clock. 
This allows us to decompose the Hilbert space on which $H$ acts as 
$\mathcal{H} = ( \mathcal{H}_{work} \otimes \mathcal{H}_{legal} )
			\oplus ( \mathcal{H}_{work} \otimes  \mathcal{H}^{\perp}_{legal} ).$
The subspace of legal clock states $\mathcal{H}_{legal}$ depends on the specific realization of the 
clock register.
The other terms in the Hamiltonian also depend on the realization of the clock. However, their restriction to 
$\mathcal{H}_{work} \otimes \mathcal{H}_{legal}$ has a simple form.
The term $H_{init}$ penalizes computations that are not initialized properly at the first clock time.
Restricted to the legal clock subspace, it reads:
\begin{eqnarray} \label{restinit}
H'_{init} &=&  \sum_{n \in ancilla}\ket{1}\bra{1}_n \otimes \ket{C_1}\bra{C_1}_{clock}
\end{eqnarray}
The term $H_{out}$ penalizes computations that do not output ``yes'' at the final clock time. Its restriction
to the legal clock space is:
\begin{eqnarray} \label{restout}
H'_{out} &=& \ket{0}\bra{0}_{out} \otimes \ket{C_{L+1}}\bra{C_{L+1}}_{clock}.
\end{eqnarray}
Finally, $H_{prop}^{(k)}$ 
verifies that the state properly encodes the computation. 
It penalizes all states for which
the components with the clock register in times $k$ and $k+1$, i.e. 
$\ket{\phi_k}_{work}\ket{k}_{clock}$ and $\ket{\phi_{k+1}}_{work}\ket{k+1}_{clock}$, 
are not related by $\ket{\phi_{k+1}} = U_k\ket{\phi_k}$.
The restriction of $H_{prop}^{(k)}$ to $\mathcal{H}_{work} \otimes \mathcal{H}_{legal}$ is:
\begin{eqnarray} \label{restprop}
H'^{(k)}_{prop} &=& \half \left( \ii_{work} \otimes \ket{C_k}\bra{C_k} 
		+ \ii_{work} \otimes \ket{C_{k+1}}\bra{C_{k+1}} 
		- U_k \otimes \ket{C_{k+1}}\bra{C_k}  
		- U_k^{\dagger} \otimes \ket{C_k}\bra{C_{k+1}} \right).
\end{eqnarray}
The idea is that if $U$ is a verifier circuit that outputs ``yes'' on $\ket{\varphi}$ with high probability, 
then the state
\begin{eqnarray}
  \ket{\psi} = \sum_{k=1}^{L+1} U_{k-1} \cdots U_2 U_1 
  	\Big( \ket{\varphi}_{input} \otimes \ket{0^{\otimes N_a}}_{ancilla} \Big) \otimes \ket{C_k}_{clock} 
  		\equiv \sum_{k=1}^{L+1} \ket{\phi_k}_{work} \otimes \ket{C_k}_{clock} \label{history}
\end{eqnarray}
which encodes the history of the computation of $U$ on $\ket{\varphi} \otimes \ket{0\dots0}$, is the ground
state of $H$ and has a small eigenvalue. On the other hand, if $U$ is a verifier circuit that outputs ``no'' 
with high probability, then any state will have high energy, either because it does not encode a legal 
computation, or because the legal computation it encodes is not likely to output ``yes''.

Using this idea it was shown by Kitaev in \cite{Kitaevbook} that 5-local Hamiltonian
is QMA complete. Kitaev's construction uses a unary clock on $L+1$ clock qubits with $L+1$ legal clock states
$\ket{C_k}_{clock} = \ket{1_1\dots 1_k 0_{k+1} \dots 0_{L+1}}$.
It is simple to advance this clock by just flipping the $(k+1)$-th clock qubit. 
However, to recognize that the clock is in the state $\ket{k}$, we need to look
at the two clock qubits $(k,k+1)$ and verify that they are in the state $\ket{10}$.
To check whether a state properly encodes a computational step
$k \rightarrow k+1$, one needs to compare the work qubits at clock states $\ket{k}$ and $\ket{k+1}$.
Therefore, the terms $H^{(k)}_{prop}$ that verify that a state
properly encodes a computation, must couple 5 qubits; 
the 3 clock qubits $(k,k+1,k+2)$ needed to recognize $\ket{k}, \ket{k+1}$ and the two work qubits on which 
the 2-qubit gate $U_k$ acts. Therefore, this realization is 5-local.

In \cite{KR03} it was shown that 3-local Hamiltonian is QMA complete.
This result was further improved to show that 2-local Hamiltonian is QMA complete \cite{KKR04}.
Both constructions use the same unary clock realization described above. Since the terms in the Hamiltonian are
no longer 5-local, the corresponding terms in $H_{prop}^k$ do not only verify proper application of
$U_k$, but also induce transitions from legal clock states into illegal ones (the subspace
$\mathcal{H}_{work} \otimes \mathcal{H}_{legal}$ is no longer invariant under the action of $H$). To fix this, the penalty
associated with illegal clock states is made high (scaling as poly$(N,L)$), effectively forcing the
ground state of the Hamiltonian to reside in the subspace of legal clock states.
This increases one of the energy scales of the problem to $\norm{H}=O(L^2)$,
because there are $O(L)$ terms with weights that scale as $O(L)$.
The new construction we describe in section \ref{3loc} shows that the 3-local Hamiltonian problem is QMA complete 
using only terms with constant operator norms, with the norm of our Hamiltonian scaling as $\norm{H}=O(L)$.


\subsection{Quantum k-SAT}
\label{QKSsection}

The quantum $k$-SAT promise problem was introduced by Bravyi \cite{Bravyi} as an analogue
of classical $k$-SAT. The problem is to determine whether the Hamiltonian $H_{qks}$ acting on the space 
of $N$ qubits has a zero eigenvalue, or whether all its eigenvalues are higher than 
$\epsilon = \Omega(1/\textrm{poly}(N))$. Also,
\begin{eqnarray}
	H_{qks} = \sum P_i, \label{QKS}
\end{eqnarray}
where each $P_i = P_i^2 =  \ii^{\otimes(N-k)} \otimes  \ket{\psi_i}\bra{\psi_i}_{\{q^{i}_1\dots q^{i}_k\}}$ 
is a projector acting nontrivially on $k$ qubits $\{q^{i}_1\dots q^{i}_k\}$. 
If all the projectors $P_i$ commute, we can transform the states $\ket{\psi_i}$ into computational
basis states such as $\ket{001}$ and retrieve classical k-SAT with $\epsilon=1$.

In \cite{Bravyi} it was proved that quantum $k$-SAT belongs to QMA$_1$ for any constant $k$.
It was further shown that quantum 4-SAT is QMA$_1$ complete using a new realization of the clock.
Bravyi uses $L+1$ clock particles with 4 states: unborn, active 1 ($a_1$, input for a gate), 
active 2 ($a_2$, output of a gate), and dead.
These 4 states of a clock particle are easily realized by two qubits per clock particle. 
There are $2L$ legal clock states: 
\begin{eqnarray}
	\ket{C_{2k-1}} = \kets{\underbrace{d\dots d}_{k-1} a_1 \underbrace{u \dots u}_{L-k}},
	\qquad \textrm{and} \qquad 	
	\ket{C_{2k}} = \kets{\underbrace{d \dots d}_{k-1} a_2 \underbrace{u \dots u}_{L-k} },
\end{eqnarray}
for $1\leq k \leq L$. A clock Hamiltonian $H_{clock} = H_{clockinit} + \sum_{k=1}^{L-1} H^{(k)}_{clock}$ 
is required to check whether the states of the clock are legal.
\begin{eqnarray}
	H^{(k)}_{clock} &=& \ket{d}\bra{d}_k \otimes \ket{u}\bra{u}_{k+1} \\ 
				&+& \ket{u}\bra{u}_k \otimes \Big(\ket{d}\bra{d}  + 
				  \ket{a_1}\bra{a_1} + \ket{a_2}\bra{a_2} \Big)_{k+1} \nonumber \\
		&+& \Big( \ket{a_1}\bra{a_1} + \ket{a_2}\bra{a_2} \Big)_k \otimes 
					\Big( \ket{a_1}\bra{a_1} + \ket{a_2}\bra{a_2} + \ket{d}\bra{d} \Big)_{k+1}, \nonumber \\
	H_{clockinit} &=& \ket{u}\bra{u}_1 + \ket{d}\bra{d}_L. \label{clockinitbravyi}
\end{eqnarray}
The Hamiltonian checking the correct application of gates is $H_{prop} = \sum_{k=1}^{L} H_{prop}^{(k)}$, with
\begin{eqnarray}
	H_{prop}^{(k)} = \frac{1}{2} \Big( \ii \otimes \ket{a_1}\bra{a_1}_k + \ii \otimes \ket{a_2}\bra{a_2}_k 
	- U_k \otimes \ket{a_2}\bra{a_1}_k - U^{\dagger}_k \otimes \ket{a_1}\bra{a_2}_k \Big). \label{prop4}
\end{eqnarray}
Each such term verifies the correct application of the gate $U_k$ 
between the states $\ket{a_1}$ and $\ket{a_2}$ of the $k$-th clock particle.
This only requires interactions of the $k$-th clock particle (qubit pair) and the two work 
qubits the gate $U_k$ is applied to. Each of the terms is thus a 4-local projector. 

We need another Hamiltonian term to propagate the clock state $\ket{C_{2k}}$ into $\ket{C_{2k+1}}$ 
while leaving the work qubits untouched (that is, for the ground state 
$\ket{\psi_{2k}}_{work} = \ket {\psi_{2k+1}}_{work}$). This is done by the 4-local clock-propagation Hamiltonian 
$H_{clockprop} = \sum_{k=1}^{L-1} H_{clockprop}^{(k)}$, with
\begin{eqnarray}
	H_{clockprop}^{(k)} &=& 
		\frac{1}{2} \Big( \ket{a_2}\bra{a_2}_k \otimes \ket{u}\bra{u}_{k+1} 
					+ \ket{d}\bra{d}_k \otimes \ket{a_1}\bra{a_1}_{k+1} \Big) \\
		&-& \frac{1}{2} \Big( \ket{d}\bra{a_2}_k \otimes \ket{a_1}\bra{u}_{k+1} 
			+ \ket{a_2}\bra{d}_k \otimes \ket{u}\bra{a_1}_{k+1} \Big). \nonumber
\end{eqnarray}
The final ingredients in this construction are
\begin{eqnarray}
	H_{init} &=&  \sum_{n=1}^{N_a}\ket{1}\bra{1}_n \otimes \ket{a_1}\bra{a_1}_1, \\
	H_{out} &=&  \ket{0}\bra{0}_{out} \otimes \ket{a_2}\bra{a_2}_L.
\end{eqnarray}
Applying Kitaev's methods \cite{Kitaevbook} to this construction, Bravyi shows that the quantum 4-SAT Hamiltonian (a sum of 4-local projectors)
\begin{eqnarray}
	H &=& H_{clock} + H_{clockprop} + H_{init} + H_{out} + H_{prop} \label{ham4}
\end{eqnarray} 
is QMA$_1$ complete.

Bravyi's original definition required all of the terms in the Hamiltonian to be projectors.
However, using $k$-local positive operator terms $H^{+}_i$ with zero ground state 
and constant norm instead of projectors $P_i$ in \eqref{QKS} is an equivalent problem.
Quantum $k$-SAT with positive operators contains quantum $k$-SAT with projectors.
On the other hand, if one is able to solve quantum $k$-SAT with projectors, one can
solve quantum $k$-SAT with positive operators as well. 
For each positive operator $H^{+}_i$, we define a projector $P^{+}_i$ with the same ground state subspace. 
If $H_P = \sum P^{+}_i$ has a zero ground state, so does $H_{+}=\sum H^{+}_i$.
If the ground state energy of $H_P = \sum P^{+}_i$ is greater than $\ep$, 
the ground state energy of $H_{+}$ is greater than $c\ep$, where $c$ is a constant. We are thus allowed
to use positive operator terms in our Hamiltonians instead of just restricting ourselves to projectors.


\section{A qutrit clock implementation} \label{3sat}

In this section we present a new realization of the clock which builds on Bravyi's quantum 4-SAT
realization described above. Using this clock construction, 
we prove that quantum 3-SAT for qutrits is QMA$_1$-complete. 

First, we need to show that quantum 3-SAT with qutrits is in QMA. We can use
Bravyi's proof that quantum $k$-SAT
for qubits is in QMA$_1$ for any constant $k$. 
Given an instance of quantum 3-SAT for qutrits, we convert it into an instance of
quantum 6-SAT for qubits by encoding each qutrit in two qubits and projecting out one of the
four states. According to Bravyi, this problem is in QMA$_1$ and therefore so is the original
quantum 3-SAT problem with qutrits.

For the other direction in the proof, we 
need to construct a quantum 3-SAT Hamiltonian for qutrits, corresponding
to a given quantum verifier circuit $U$ for a problem in QMA. 
The terms in the Hamiltonian we will construct act on the space of one qutrit 
and two qubits (particles with dimensions $3\times 2\times 2$). 

\subsection{Clock register construction}

The clock-register construction in the previous section required 
4 states for each clock particle: $\ket{u},\ket{a_1},\ket{a_2}$ and $\ket{d}$. 
Let us first understand why Bravyi's construction requires two ``inactive'' states: $\ket{d}$ and $\ket{u}$.
If we only use $\ket{d}$ (i.e., have legal 
clock states of the form $\ket{d\dots d a_1 d \dots d}$ and $\ket{d\dots d a_2 d \dots d}$), we immediately get a 3-local 
Hamiltonian for qutrits. However, in Bravyi's construction, the first clock particle is never
in the state $\ket{u}$, and the last one is never in the state $\ket{d}$ (see \eqref{clockinitbravyi}). This ensures that at least one clock
particle is in an active state. When not using the state $\ket{u}$, we can no longer 
exclude the state with no active particles $\ket{dd\dots d}$ in a simple local fashion.

\begin{figure}[h]
	\includegraphics[width=3.5in]{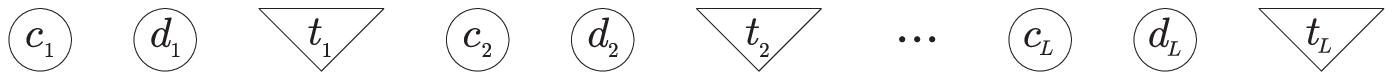} 
	\caption{Clock register consisting of $2L$ qubits and $L$ qutrits. \label{construct3d}}
\end{figure}
We fix this by modifying the clock register as shown in in Fig.\ref{construct3d}.
The clock register now consists of $2L$ qubits and $L$ qutrits.
The $2L$ qubits $c_1,d_1,\dots,c_k,d_k$ play the role of the 
usual unary $\ket{1\dots 11 00 \dots0}$ clock representation, 
while the $L$ qutrits $t_1,t_2, \dots t_k$ play the role of Braviy's clock with just three states
($d,a_1,a_2$).

We define the legal clock space $\mathcal{H}_{legal}$ as the space spanned by the $3L$ states $\ket{C_m}$.
These states are defined for $1\leq k \leq L$ as follows:
\begin{eqnarray}
	\ket{C_{3k-2}} &=& \kets{\underbrace{(11d)(11d)\dots(11d)}_{k-1\,\,\textrm{times}} (10d) 
		\underbrace{(00d)(00d)\dots(00d)}_{L-k \,\,\textrm{times}}}, \label{qutritlegal} \\
	\ket{C_{3k-1}} &=& \kets{\underbrace{(11d)(11d)\dots(11d)}_{k-1\,\,\textrm{times}} (11a_1) 
		\underbrace{(00d)(00d)\dots(00d)}_{L-k \,\,\textrm{times}}}, \nonumber \\
	\ket{C_{3k}} &=& \kets{\underbrace{(11d)(11d)\dots(11d)}_{k-1\,\,\textrm{times}} (11a_2) 
		\underbrace{(00d)(00d)\dots(00d)}_{L-k \,\,\textrm{times}}}. \nonumber
\end{eqnarray}
The first state ($\ket{C_{3k-2}}$) corresponds to the time when the
qubits are ``in transport'' from the the previous gate to the to the current ($k$th) gate.
The second one corresponds to the time right before application of gate $U_k$ 
and the third corresponds to the time right after the gate $U_k$ was applied.
The structure of such clock register can be understood as two coupled
``unary'' clocks, the qubit one ($c_k,d_k$) of the $11\dots11100\dots00$ type and the qutrit one 
(the $t_k$'s) of the $00\dots00100\dots00$ type. 
Formally, the legal clock states satisfy the following constraints:
\begin{enumerate}
\item if $d_k$ is 1, then $c_k$ is 1.
\item if $c_{k+1}$ is 1, then $d_k$ is 1.
\item if $t_k$ is active ($a_1/a_2$), then $d_k$ is 1.
\item if $t_k$ is active ($a_1/a_2$), then $c_{k+1}$ is 0.
\item if $d_k$ is 1 and $c_{k+1}$ is 0, then $t_{k}$ is not dead $(d)$.
\item $c_1$ is 1.
\item if $d_L$ is 1, then $t_{L}$ is not dead.
\end{enumerate}
The last two conditions are required to exclude the clock states $\ket{(00d)(00d)\dots (00d)}$ 
and $\ket{(11d)(11d) \dots(11d)}$ that have no active clock terms.
The clock Hamiltonian $H_{clock} = H_{clockinit} + \sum_{k=1}^{L} H_{clock1}^{(k)} + \sum_{k=1}^{L-1} H_{clock2}^{(k)}$ 
verifies the above constraints.
\begin{eqnarray}
		H_{clock1}^{(k)} &=& 
				\ket{01}\bra{01}_{c_k,d_k} 
				+ \ket{0}\bra{0}_{d_k} \otimes \Big( \ket{a_1}\bra{a_1} + \ket{a_2}\bra{a_2} \Big)_{t_k}, \label{clock3} \\
		H_{clock2}^{(k)} &=& 
				\ket{01}\bra{01}_{d_k,c_{k+1}} 
				+ \Big( \ket{a_1}\bra{a_1} + \ket{a_2}\bra{a_2} \Big)_{t_k} \otimes \ket{1}\bra{1}_{c_{k+1}} \nonumber \\
			&+& \ket{1d0}\bra{1d0}_{d_k,t_k,c_{k+1}}, \nonumber \\
		H_{clockinit} &=&
				\ket{0}\bra{0}_{c_1} 
			+ \ket{1}\bra{1}_{d_{L}} \otimes \ket{d}\bra{d}_{t_{L}}. \nonumber
\end{eqnarray}
Only the last term in $H_{clock2}^{(k)}$ is a 3-local projector, 
acting on the space of two qubits and one qutrit. 
The rest of the terms are 2-local projectors on two qubits, or a qubit and a qutrit.
The space of legal clock states $\mathcal{H}_{legal}$
is the kernel of the clock Hamiltonian $H_{clock}$.

\subsection{Checking correct application of gates and clock propagation}

The gate-checking Hamiltonian $H_{prop} = \sum_{k=1}^{L} H_{prop}^{(k)}$ is an analogue of \eqref{prop4}, with
\begin{eqnarray}
	H_{prop}^{(k)} = \frac{1}{2} \Big( \ii_{work} \otimes \ket{a_1}\bra{a_1}_{t_k} 
		+ \ii_{work} \otimes \ket{a_2}\bra{a_2}_{t_k}
	- U_k \otimes \ket{a_2}\bra{a_1}_{t_k} - U^{\dagger}_k \otimes \ket{a_1}\bra{a_2}_{t_k} \Big). \label{prop3}
\end{eqnarray}

\begin{figure}[h]
	\includegraphics[width=3.6in]{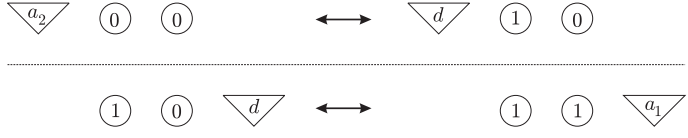} 
	\caption{Illustration of the two-step clock pointer propagation. \label{clockprop3d}}
\end{figure}
The clock propagation proceeds in two steps. First, the ``active'' spot in the clock register moves from
the state $\ket{a_2}$ of the qutrit $t_k$ to the $\ket{10}$ state of the next two qubits $c_{k+1},d_{k+1}$.
After this, it moves into the state $\ket{a_1}$ of the next qutrit $t_{k+1}$, as in Fig.\ref{clockprop3d}.
The Hamiltonian checking whether this happened, while the work qubits were left untouched, is 
$H_{clockprop} = \sum_{k=1}^{L} H_{clockprop1}^{(k)} + \sum_{k=1}^{L-1} H_{clockprop2}^{(k)}$, with
\begin{eqnarray}
	H_{clockprop1}^{(k)} &=& \frac{1}{2} \Big( 
				\ket{10}\bra{10}_{c_k,d_k} \otimes \ket{d}\bra{d}_{t_{k}} 
			+ \ket{11}\bra{11}_{c_k,d_k} \otimes \ket{a_1}\bra{a_1}_{t_{k}}
				 	\Big) \label{clockprop3} \\
	&-& \frac{1}{2} \Big( 
				\ket{11}\bra{10}_{c_k,d_{k}} \otimes \ket{a_1}\bra{d}_{t_{k}} 
			+ \ket{10}\bra{11}_{c_k,d_{k}} \otimes \ket{d}\bra{a_1}_{t_{k}} 
					\Big), \nonumber \\
	H_{clockprop2}^{(k)} &=& \frac{1}{2} \Big( 
				 \ket{a_2}\bra{a_2}_{t_{k}} \otimes \ket{00}\bra{00}_{c_{k+1},d_{k+1}} 
			+	 \ket{d}\bra{d}_{t_{k}} \otimes \ket{10}\bra{10}_{c_{k+1},d_{k+1}} 
				 	\Big) \nonumber \\
	&-& \frac{1}{2} \Big( 
				 \ket{d}\bra{a_2}_{t_{k}} \otimes \ket{10}\bra{00}_{c_{k+1},d_{k+1}} 
			+	 \ket{a_2}\bra{d}_{t_{k}} \otimes \ket{00}\bra{10}_{c_{k+1},d_{k+1}} 
					\Big). \nonumber
		\end{eqnarray}
The input Hamiltonian checks whether the computation has properly initialized ancilla qubits.
\begin{eqnarray}
	H_{init} &=& \sum_{n=1}^{N_a} \ket{1}\bra{1}_n \otimes \ket{a_1}\bra{a_1}_{t_1}. \label{init3} 
\end{eqnarray}
Finally, the output Hamiltonian checks whether the result of the computation was 1.
\begin{eqnarray}
	H_{out} &=& \ket{0}\bra{0}_{out} \otimes \ket{a_2}\bra{a_2}_{t_{L}}. \label{out3} 
\end{eqnarray}

All of the terms coming from \eqref{clock3} -- \eqref{out3} in the Hamiltonian 
\begin{eqnarray}
	H &=& H_{clock} + H_{clockprop} + H_{init} + H_{out} + H_{prop}. \label{ham3}
\end{eqnarray}
are projectors. Therefore, 
the ground state has energy zero if and only if there exists 
a zero energy eigenstate of all of the terms.
If there exists a witness $\ket{\varphi}$ 
on which the computation $U$ gives the result 1 with probability 1,
we can construct a computational history state \eqref{history} 
for a modified circuit $\tilde{U}= U_L \cdot \ii \cdot \ii \cdot U_{L-1} \cdot \ii\cdot \ii \cdots U_1 \cdot \ii$,
where the ``identity'' gates correspond to the clock propagation in our construction, with nothing
happening to the work qubits.
This state is a zero eigenvector of all of the terms in the Hamiltonian \eqref{ham3}.

We now need to prove that if no witness exists (the
answer to the problem is ``no''), then the ground state energy of \eqref{ham3} is
 $\Omega(1/poly(N,L))$. 
Let us decompose the Hilbert space into 
\begin{eqnarray}
	\mathcal{H} = \left( \mathcal{H}_{work} \otimes \mathcal{H}_{legal} \right)
			\oplus \left( \mathcal{H}_{work} \otimes  \mathcal{H}^{\perp}_{legal} \right).
\end{eqnarray}
where $\mathcal{H}_{legal}$ is the space of legal clock states (on which $H_{clock}\ket{\alpha} =0$).
The Hamiltonian \eqref{ham3} leaves this decomposition invariant, because it does not
induce transitions between legal and illegal clock states.
Since any state in $\mathcal{H}_{legal}^{\perp}$ violates at least one
term in $H_{clock}$, the lowest eigenvalue of the 
restriction of \eqref{ham3} to $\mathcal{H}_{work}\otimes \mathcal{H}_{legal}^{\perp}$
is at least 1.
On the other hand, the restriction of $H$ to the legal clock space 
is identical to the legal clock space restriction of Bravyi's Hamiltonian \eqref{ham4} 
from the previous section. Therefore, his proof using the methods
of Kitaev \cite{Kitaevbook} applies to our case as well. He shows that
if a no witness state for the quantum circuit $U$ exists, then
the ground state energy of the restriction of \eqref{ham4} to $\mathcal{H}_{work} \otimes \mathcal{H}_{legal}$
is $\Omega(1/poly(N,L))$.
This means that if there is no witness state for the verifier circuit $U$, the ground state 
of \eqref{ham3} is $\Omega(1/poly(N,L))$. This concludes the proof
that quantum 3-SAT with qutrits (in fact, quantum 3-SAT on particles with dimensions 
$3\times 2\times 2$, a qutrit and two qubits)
is QMA$_1$ complete.

The existence of another $3 \times 2 \times 2$ construction for quantum 3-SAT
(i.e., a Hamiltonian with terms acting on one qutrit and two qubits)
was already mentioned in \cite{Bravyi} as \cite{DiVincenzo}, though that construction was not specified. 
We think it is instructive to write our
result explicitly since it serves as a natural intermediate step towards the new 3-local 
Hamiltonian construction described in the following section.


\section{The new 3-local QMA complete construction (for qubits)} \label{3loc}

\subsection{Clock register construction}
\begin{figure}[h]
	\includegraphics[width=3.6in]{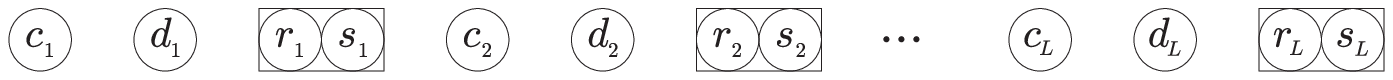}
	\caption{Clock register construction with $2L+2L$ qubits. \label{construct3loc}}
\end{figure}

In Bravyi's Quantum 4-SAT construction \cite{Bravyi}, the clock particles (qubit pairs) 
can be in 4 states. In the previous section, we required only 3 states of the clock particles
and used qutrits as particles with these three states.
We start with the clock-register construction (see Fig.\ref{construct3d}) from the previous section,
with legal states as in \eqref{qutritlegal}. However, we now encode the three states 
of every clock qutrit $t_k$ using a pair of qubits $r_k, s_k$. The new clock register is depicted 
in Fig.\ref{construct3loc}.
\begin{eqnarray}
	\ket{a_1}_{t_k} &\rightarrow& \frac{1}{\sqrt{2}}\left(\ket{01}-\ket{10}\right)_{r_k,s_k}, \qquad 
			\ket{d}_{t_k} \rightarrow \ket{00}_{r_k,s_k}, \\
	\ket{a_2}_{t_k} &\rightarrow& \frac{1}{\sqrt{2}}\left(\ket{01}+\ket{10}\right)_{r_k,s_k}. \nonumber
\end{eqnarray}
This encoding allows us to construct a new 3-local Hamiltonian construction.
This Hamiltonian is be a quantum 4-SAT Hamiltonian 
whose 4-local positive operator terms consist of just 3-local interactions.

We are looking for 3-local terms that flip between the clock states 
$\ket{a_1} \leftrightarrow \ket{a_2}$, while simultaneously (un)applying a 2-qubit gate $U_k$
on two work qubits. 
We encode the active states of a clock particle into
entangled states, and thus we are able to flip between these clock states
with a term like $Z_1$, involving only one of the clock particles.
Thus our 2-qubit gate checking Hamiltonian involves only 3-local terms 
(acting on one clock qubit $r_k$ or $s_k$ and the two work qubits on which the gate $U_k$ acts).

First, we define the legal clock space $\mathcal{H}_{legal}$ as the space spanned by the 
$3L$ states $\ket{C_m}$. These states are defined for $1\leq k \leq L$ as follows (compare to \eqref{qutritlegal}):
\begin{eqnarray}
	\ket{C_{3k-2}} &=& \kets{\underbrace{(11)(00)\dots (11)(00)}_{k-1\,\,\textrm{times}}}
		\otimes \ket{10}_{c_k,d_k} \otimes \ket{00}_{r_k, s_k} \otimes 
		\kets{\underbrace{(00)(00)\dots (00)(00)}_{L-k \,\,\textrm{times}}}, \label{threeloclegal} \\
	\ket{C_{3k-1}} &=& 
		\kets{\underbrace{(11)(00)\dots (11)(00)}_{k-1\,\,\textrm{times}}}
		\otimes \kets{11}_{c_k,d_k} \otimes \frac{1}{\sqrt{2}}\left(\ket{01}-\ket{10}\right)_{r_k,s_k}
		\otimes \kets{\underbrace{(00)(00)\dots (00)(00)}_{L-k \,\,\textrm{times}}}, \nonumber \\
	\ket{C_{3k}} &=& 
		\kets{\underbrace{(11)(00)\dots (11)(00)}_{k-1\,\,\textrm{times}}}
		\otimes \kets{11}_{c_k,d_k} \otimes \frac{1}{\sqrt{2}}\left(\ket{01}+\ket{10}\right)_{r_k,s_k}
		\otimes \kets{\underbrace{(00)(00)\dots (00)(00)}_{L-k \,\,\textrm{times}}}. \nonumber
\end{eqnarray}
Similarly to the construction of the previous section, the first state ($\ket{C_{3k-2}}$) corresponds to the time when the
qubits are ``in transport'' from the the previous gate to the to the current ($k$th) gate.
The second one corresponds to the time right before application of gate $U_k$ 
and the third corresponds to the time right after the gate $U_k$ was applied.

Formally, the legal clock states for this construction satisfy the following constraints:
\begin{enumerate}
\item if $d_k$ is 1, then $c_k$ is 1.
\item if $c_{k+1}$ is 1, then $d_k$ is 1.
\item the pair $r_k, s_k$ is not in the state $\ket{11}$.
\item if the pair $r_k, s_k$ is active (in the state $(\ket{01}\pm \ket{10})/\sqrt{2}$), then $d_k$ is 1.
\item if the pair $r_k, s_k$ is active (in the state $(\ket{01}\pm \ket{10})/\sqrt{2}$), then $c_{k+1}$ is 0.
\item if $d_k$ is 1 and $c_{k+1}$ is 0, then the pair $r_k, s_k$ is not dead (in the state $\ket{00}$).
\item $c_1$ is 1.
\item if $d_L$ is 1, then the pair $r_L,s_L$ is not dead (in the state $\ket{00}$).
\end{enumerate}
The last two conditions are required to make the clock states $\ket{(00)(00)\dots (00)(00)}$ 
and $\ket{(11)(00)\dots(11)(00)}$ with no active spots illegal.
The clock Hamiltonian $H_{clock} = H_{clockinit} + \sum_{k=1}^{L} H_{clock1}^{(k)} + \sum_{k=1}^{L-1} H_{clock2}^{(k)}$ 
verifies the above constraints.
\begin{eqnarray}
		H_{clock1}^{(k)} &=& 
				\ket{01}\bra{01}_{c_k,d_k} 
				+ \ket{0}\bra{0}_{d_k} \otimes \big( \ket{1}\bra{1}_{r_k} +\ket{1}\bra{1}_{s_k} \big)
				+ \ket{11}\bra{11}_{r_k,s_k}, \label{clock4} \\
		H_{clock2}^{(k)} &=& 
				\ket{01}\bra{01}_{d_k,c_{k+1}} 
				+ \big( \ket{1}\bra{1}_{r_k} +\ket{1}\bra{1}_{s_k} \big) \otimes \ket{1}\bra{1}_{c_{k+1}} + h_4^{(k)}, \nonumber \\
		h_4^{(k)} &=& 
				\ket{1}\bra{1}_{d_k} \otimes \frac{1}{2}\,(Z_{r_k}+Z_{s_k}) \otimes \ket{0}\bra{0}_{c_{k+1}} 
				+ \ket{11}\bra{11}_{r_k,s_k}, \nonumber \\
		H_{clockinit} &=&
				\ket{0}\bra{0}_{c_1} 
			+ \ket{1}\bra{1}_{d_{L}} \otimes \ket{00}\bra{00}_{r_L,s_L}. \nonumber
\end{eqnarray}
All of the terms involve only 3-local interactions.
All terms in $H_{clock1}^{(k)}$, $H_{clock2}^{(k)}$ and $H_{clockinit}$, are are projectors. 
The term $h_4^{(k)}$ corresponds to the sixth legal state condition.
It is a 4-local projector onto the space spanned by (illegal clock) states $\ket{1_{d_k}(00)_{r_k,s_k}0_{c_{k+1}}}$, 
$\ket{0(11)0}$, $\ket{0(11)1}$ and $\ket{1(11)1}$.
Note that even though $h_4^{(k)}$ is a 4-local projector, it is only constructed of 3-local terms.

\subsection{Checking gate application with 3-local terms}
Let us start by writing out a Hamiltonian that checks the correct application of a single-qubit gate $U_k$.
\begin{eqnarray}
	H_{prop}^{(k),\,one-qubit} = \frac{1}{2}\left( 
		\begin{array}{rr}
			\ii\otimes \ket{01-10}\bra{01-10}_{r_k, s_k} 
				- & U_k \otimes \ket{01+10}\bra{01-10}_{r_k, s_k}\\
			\ii\otimes \ket{01+10}\bra{01+10}_{r_k, s_k}
				- & U_k^{\dagger} \otimes \ket{01-10}\bra{01+10}_{r_k, s_k}
			\end{array}
		\right),
\end{eqnarray}
where $\ket{01\pm 10}$ is a shortcut notation for the normalized entangled states $(\ket{01}\pm \ket{10})/\sqrt{2}$.
This Hamiltonian is a 3-local projector. 
Note that in the case $U_k=\ii$, this Hamiltonian becomes the projector $(\ii-X)/2$
on the space of active clock states $\{\ket{01-10},\ket{01+10}\}$.

For a two-qubit gate $U_k$, the above construction would be 4-local. However, we are be able to 
construct this 4-local projector using
only 3-local terms. To do this, we require the 2-qubit gate to be symmetric"' $U_k=U_k^{\dagger}$. 
This is a universal construction, 
since the symmetric gate \textsc{CNOT} (or \textsc{C}$_{\phi}$) is universal.
Now we can write 
\begin{eqnarray}
	H^{(k),\,two-qubit}_{prop} 
		= \frac{1}{2}\Big( \ii \otimes \frac{1}{2}\,( \ii - Z_{r_k} Z_{s_k} ) -  U_k \otimes  \frac{1}{2}\,(Z_{r_k}-Z_{s_k}) \Big).
\end{eqnarray}
The first term in this Hamiltonian, $( \ii - Z_{r_k} Z_{s_k} )/2$, is a projector onto the space of active clock states,
$\ket{01\pm10}_{r_k,s_k}$, as we needed. The second term contains $(Z_{r_k}-Z_{s_k})/2$, which has zero eigenvalues
for the states $\ket{00}_{r_k,s_k}$ and $\ket{11}_{r_k,s_k}$, and flips between the states 
$\ket{01-10}_{r_k,s_k}\leftrightarrow\ket{01+10}_{r_k,s_k}$.
Altogether, this is a 4-local projector made out of only 3-local terms. 

\subsection{Clock propagation}
\label{clockpropagsection}

\begin{figure}[ht]
	\includegraphics[width=3.6in]{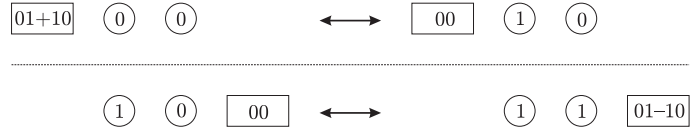}
	\caption{Illustration of the two-step clock pointer propagation. \label{figclockprop}}
\end{figure}

After a gate $U_k$ is applied, we need to ``propagate'' the pointer (the active state of the qubit pair
$r_k,s_k$) to the next pair of qubits $r_{k+1},s_{k+1}$.
This is done in two steps, as shown in Fig.\ref{figclockprop}.

For each step, we want to write a 3-local positive Hamiltonian 
with terms acting on 4 consecutive qubits $r_k,s_k,c_{k+1},d_{k+1}$ 
(for the second step of the clock pointer propagation, the four qubits in play are $c_k,d_k,r_k,s_k$), 
with zero eigenvalue for the legal clock-propagation states, and perhaps
also some illegal clock states, which will be disallowed by other terms in the Hamiltonian ($H_{clock}$).
For the first step, these desired eigenvectors with zero eigenvalues are 
\begin{eqnarray}
	\ket{\alpha_{1}}_{r_k,s_k,c_{k+1},d_{k+1}} &=& \ket{00}_{r_k,s_k} \ket{00}_{c_{k+1},d_{k+1}}, \\
	\ket{\alpha_{2}}_{r_k,s_k,c_{k+1},d_{k+1}} &=& \frac{1}{\sqrt{2}}(\ket{01}-\ket{10})_{r_k,s_k} \ket{00}_{c_{k+1},d_{k+1}}, \nonumber \\
	\ket{\alpha_{3}}_{r_k,s_k,c_{k+1},d_{k+1}} &=& \frac{1}{2}(\ket{01}+\ket{10})_{r_k,s_k} \ket{00}_{c_{k+1},d_k+1}
				 																 + \frac{1}{\sqrt{2}}\ket{00}_{r_k,s_k} \ket{10}_{c_{k+1},d_{k+1}}, \nonumber\\
	\ket{\alpha_{4}}_{r_k,s_k,c_{k+1},d_{k+1}} &=& \ket{00}_{r_k,s_k} \ket{11}_{c_{k+1},d_{k+1}}. \nonumber
\end{eqnarray}
The state that we want to exclude (make it a nonzero eigenvector) is the legal clock state with incorrect pointer propagation:
\begin{eqnarray} \label{exclude1}
	\ket{\alpha^{\perp}}_{r_k,\dots,d_{k+1}} &=& 
			\frac{1}{2}(\ket{01}+\ket{10})_{r_k,s_k} \ket{00}_{c_{k+1},d_k+1}
		- \frac{1}{\sqrt{2}}\ket{00}_{r_k,s_k} \ket{10}_{c_{k+1},d_{k+1}}.
\end{eqnarray}
Let us present such a Hamiltonian for this step.
\begin{eqnarray}
	H^{(k)}_{clockprop1} &=& \left(
			\begin{array}{lll}
				&\ket{10} \bra{10}_{c_{k+1},d_{k+1}}  
					& - \frac{1}{\sqrt{2}}\Big( \ket{0}\bra{1}_{r_k}+\ket{0}\bra{1}_{s_k} \Big) 
							\otimes \ket{10}\bra{00}_{c_{k+1},d_{k+1}} \\
				+ & \frac{1}{2}\left(\ket{01}+\ket{10}\right)\left(\bra{01}+\bra{10}\right)_{r_k,s_k} \quad							
					& - \frac{1}{\sqrt{2}}\Big( \ket{1}\bra{0}_{r_k}+\ket{1}\bra{0}_{s_k} \Big) 
							\otimes \ket{00}\bra{10}_{c_{k+1},d_{k+1}} \\
			\end{array}
		\right) \label{prop41}\\
	&+& 2\ket{11}\bra{11}_{r_k,s_k}. \nonumber
\end{eqnarray}
This is a positive operator with eigenvalues 
0 ($\times 7$), 
1 ($\times 4$), 
2 ($\times 3$) and  
3 ($\times 2$).
Its zero energy eigenvectors are $\ket{\alpha_1}$, $\ket{\alpha_2}$, $\ket{\alpha_3}$, $\ket{\alpha_4}$ expressed above, 
and three illegal clock states, 
$\ket{00}_{r_k,s_k}\ket{01}_{c_{k+1},d_{k+1}}$, 
$(\ket{01}-\ket{10})\ket{01}$ and 
$(\ket{01}-\ket{10})\ket{11}$.
As mentioned earlier, the only purpose of this Hamiltonian term is to have positive expectation values
for the legal states of the clock register \eqref{exclude1} , which do not correctly propagate the clock. 
This Hamiltonian term is a positive operator, while Bravyi's original definition of quantum $k$-SAT
requires the terms in the Hamiltonian to be projectors. However, as we have shown at the end of Section \ref{QKSsection},
quantum $k$-SAT with positive operator terms is equivalent to quantum $k$-SAT with only projector terms.

For the second step, the desired zero energy eigenvectors are
\begin{eqnarray}
	\ket{\beta_{1}}_{c_k,d_k,r_k,s_k} &=& \ket{00}_{c_k,d_k} \ket{00}_{r_k,s_k}, \\
	\ket{\beta_{2}}_{c_k,d_k,r_k,s_k} &=& \frac{1}{\sqrt{2}}\ket{10}_{c_{k},d_{k}} \ket{00}_{r_k,s_k}
				+ \ket{11}_{c_{k},d_k} \frac{1}{2}(\ket{01}-\ket{10})_{r_k,s_k}, \nonumber\\
	\ket{\beta_{3}}_{c_k,d_k,r_k,s_k} &=& \ket{11}_{c_k,d_k} \frac{1}{\sqrt{2}}(\ket{01}+\ket{10})_{r_k,s_k}, \nonumber \\
	\ket{\beta_{4}}_{c_k,d_k,r_k,s_k} &=& \ket{11}_{c_{k},d_{k}} \ket{00}_{r_k,s_k}, \nonumber
\end{eqnarray}
and the state we want to exclude is
\begin{eqnarray}
	\ket{\beta^{\perp}}_{c_k,d_k,r_k,s_k} &=& 
			\frac{1}{\sqrt{2}}\ket{10}_{c_{k},d_{k}} \ket{00}_{r_k,s_k}
				- \ket{11}_{c_{k},d_k} \frac{1}{2}(\ket{01}-\ket{10})_{r_k,s_k}.
\end{eqnarray}
The Hamiltonian with these properties is a simple analogue of \eqref{prop41}:
\begin{eqnarray}
	H^{(k)}_{clockprop2} &=& \left(
			\begin{array}{lll}
				&\ket{10} \bra{10}_{c_{k},d_{k}}  
					& - \ket{11}\bra{10}_{c_{k},d_{k}} \otimes 
						\frac{1}{\sqrt{2}}\Big( -\ket{1}\bra{0}_{r_k}+\ket{1}\bra{0}_{s_k} \Big) \\
				+ & \frac{1}{2}\left(\ket{01}-\ket{10}\right)\left(\bra{01}-\bra{10}\right)_{r_k,s_k} \quad							
					& - \ket{10}\bra{11}_{c_{k},d_{k}} \otimes
						\frac{1}{\sqrt{2}}\Big( -\ket{0}\bra{1}_{r_k}+\ket{0}\bra{1}_{s_k} \Big) \\
			\end{array}
		\right) \label{prop42}\\
	&+& 2\ket{11}\bra{11}_{r_k,s_k}. \nonumber
\end{eqnarray}
This is again a positive operator with eigenvalues 
0 ($\times 7$), 
1 ($\times 4$), 
2 ($\times 3$) and  
3 ($\times 2$).
Its zero energy eigenvectors are $\ket{\beta_1}$, $\ket{\beta_2}$, $\ket{\beta_3}$, $\ket{\beta_4}$ expressed above, 
and three illegal clock states, 
$\ket{01}_{c_{k},d_{k}}\ket{00}_{r_k,s_k}$, 
$\ket{01}(\ket{01}+\ket{10})$ and 
$\ket{00}(\ket{01}+\ket{10})$, which are penalized by $H_{clock}$.
As mentioned earlier, the only purpose of this Hamiltonian term is to have a positive expectation value
for the legal state of the clock register $\ket{\beta^{\perp}}$, which does not correctly propagate the clock. 

Just as in the previous section, the total Hamiltonian leaves the decomposition into 
$\mathcal{H}_{legal} \otimes \mathcal{H}_{illegal}$ invariant while all illegal clock states 
violate at least one term in $H_{clock}$. Again, up to a constant prefactor, 
the restriction of $H$ to the legal clock space is the same as that of the Hamiltonian in \eqref{ham4}.
The proof of the necessary separation between positive and negative instances
then follows the proof in the previous section.
This concludes the proof that quantum 4-SAT with positive operators
made out of 3-local terms is QMA$_1$ complete.


\section{Discussion and further directions} \label{conclusions}

In this work we proved that quantum 3-SAT for particles with dimensions 
$3\times 2 \times 2$ is QMA$_1$ complete. We have shown in Section \ref{QKSsection} that
quantum $k$-SAT with positive operator terms 
(not just projectors) is equivalent to quantum $k$-SAT with projector terms. 
We presented a new 3-local construction of a quantum 4-SAT Hamiltonian
with positive operator terms,
proving that quantum 4-SAT with 3-local interactions 
is QMA$_1$ complete.

The difference between the $k$-local-Hamiltonian problem and the quantum $k$-SAT problem 
is that for a ``yes'' instance of the latter, 
a common ground state of {\em all} of the terms in the Hamiltonian must exist. 
The $k$-local-Hamiltonian problem is a quantum analogue to classical MAX-$k$-SAT,
where we are interested only in the properties of the ground state of the {\em total} 
Hamiltonian (the {\em sum} of the terms).
It seems natural to define quantum MAX-$k$-SAT as the $k$-local-Hamiltonian problem 
restricted to positive operator terms.
Quantum MAX-$k$-SAT is equivalent to the $k$-local-Hamiltonian problem. Obviously, $k$-local-Hamiltonian contains
quantum MAX-$k$-SAT. On the other hand, there is a straightforward reduction from $k$-local-Hamiltonian to
quantum MAX-$k$-SAT; 
We shift the eigenvalues of each local operator $H_i$ in $k$-local-Hamiltonian 
to make it a positive operator, 
$H_i^{+} = H_i + \delta_i \ii$, appropriate for quantum MAX-$k$-SAT.
The corresponding energy parameters are $a_{MAX}=a+\sum \delta_i$ and $b_{MAX}=b+\sum \delta_i$ 
where $a$ and $b$ are parameters defined in Sec.\ref{lochamsection}.

\begin{table}[h]
\begin{tabular}{|r|l|l|}
\hline 
& \ Classical & \ Quantum \ \\
\hline 
$k$-SAT & \ $k=2$ : in P	& \ $k=2$ : in P \  \\
& \ $k\geq 3$ : NP-complete & \ $k\geq4$ : QMA$_1$-complete \ \\
\hline	
MAX-$k$-SAT & \ $k\geq2$ : NP-complete & \ $k\geq2$ : QMA-complete \ \\
\hline
\end{tabular}
\caption{Known complexity for classical and quantum SAT-type problems.}
\label{comparetable}
\end{table}

The currently known complexities of classical and quantum satisfiability problems 
are shown in Table \ref{comparetable}. 
Quantum 3-SAT contains classical 3-SAT and therefore is NP-hard. 
Unlike classical $k$-SAT, which is known to be NP-complete for $k\geq 3$, 
quantum $k$-SAT is only known to be QMA$_1$ complete for $k\geq 4$ \cite{Bravyi}. 
In our opinion, it is unlikely
that one can show that quantum 3-SAT ($k=3$) is also complete for QMA$_1$.
One indication for this arises in our numerical explorations, where random instances 
of quantum 3-SAT for a reasonable number of clauses generally have no solutions, 
unless the clauses exclude non-entangled states.
This may suggest that the hardness of quantum 3-SAT 
actually lies only in the classical instances (3-SAT)
and a classical verifier circuit for quantum 3-SAT might exist.

Another reason comes from dimension counting. This argument, however, is only valid 
for the specific encoding of a circuit into the Hamiltonian we used. 
We worked with a tensor product space
$\mathcal{H}_{work}\otimes \mathcal{H}_{clock}$, encoding the computation in the history state 
\eqref{history}.
Encoding an interaction of two qubits requires at least an $4+4=8$ dimensional space
($4$ for the two qubits before the interaction and $4$ for the qubits after the interaction).
On a first glance, a three-local projector on the space of two work qubits and one clock qubit 
($2\times2\times2=8$) seems to suffice.
However, we must ensure that this interaction only occurs at a
specific clock time. When the two work qubits interact with just a single clock qubit 
(flipping it between states 
before/after interaction) ambiguities and legal-illegal clock state 
transitions are unavoidable. This transforms the problem 
from the SAT-type (determining whether a {\em simultaneous} ground state of {\em all} terms 
in the Hamiltonian exists) 
to the MAX-SAT type problem (determining the
properties of the ground state energy of the {\em sum of terms} in the Hamiltonian). 
A single clock qubit cannot both determine the exact time of an interaction and 
distinguish between the states before and after the interaction.
We managed to overcome this obstacle by using three-dimensional clock particles with 
states $d$, $a_1$ and $a_2$ and a $2\times2\times3=12$ dimensional space for encoding the interactions. 
We believe that this can not be further improved with more clever
clock-register realizations within the $\mathcal{H}=\mathcal{H}_{work}\otimes \mathcal{H}_{clock}$ framework.

A different approach to encoding a quantum computation into the ground state of a Hamiltonian
is found in the work of Aharonov et.al \cite{AdiabaticEquivalent}. Their {\em geometric clock} idea
is to lay out the qubits in space in such a way, that the ``shape'' of the state (the shape of qubits
which are currently in active states) uniquely corresponds to a clock time. Their motivation
was to show that adiabatic quantum computation \cite{Adiabatic} is polynomially 
equivalent to the circuit model.
As a side result, they showed that the nearest-neighbor 2-local Hamiltonian problem 
with 6-dimensional particles is QMA complete. We actually know, as Kempe et al. proved, 
that even 2-local Hamiltonian with 2-dimensional particles (qubits) is QMA complete \cite{KKR04}.

We did not succeed to improve or reproduce the construction of quantum 3-SAT for particles with dimensions 
$3\times2\times2$ using the geometric clock framework. However, one
can use the idea of a geometric clock to construct quantum 2-SAT for higher dimensional particles (qudits).
We know that classical 2-SAT for particles with dimensions $3\times3$ contains graph coloring,
and is thus NP-complete. 
Using a rather straightforward modification of the Aharonov et. al. construction, one can prove 
that quantum 2-SAT for
12-dimensional particles is QMA complete. Combining the work/clock and the geometric clock constructions, 
a much tighter result can be shown. Specifically, one can construct quantum 2-SAT for particles 
with dimensions $7\times3$ and prove that 
it is QMA$_1$ complete \cite{NagajMozes2}. 
It would be interesting to find out whether these are the minimal dimensions of particles 
for which quantum 2-SAT is QMA$_1$ complete.


\section*{Acknowledgements}
We would like to thank Eddie Farhi, Seth Lloyd, Peter Shor and Jeffrey Goldstone for 
fruitful discussions. DN gratefully acknowledges support from the National Security Agency (NSA) 
and Advanced Research and Development Activity (ARDA) 
under Army Research Office (ARO) contract W911NF-04-1-0216.
SM thanks Eddie Farhi and his group at MIT for their hospitality.


\end{document}